\documentclass[aps,prb,twocolumn,showpacs,superscriptaddress]{revtex4}

\usepackage{graphicx} 

\begin{document}

\title
{Effects of hydrogen adsorption on single wall carbon nanotubes:
Metallic hydrogen decoration}

\author{O. G\"{u}lseren}
\affiliation{NIST Center for Neutron Research,
National Institute of Standards and Technology,
Gaithersburg, MD 20899}
\affiliation{Department of Materials Science and Engineering,
University of Pennsylvania, Philadelphia, PA 19104}
\author{T. Yildirim}
\affiliation{NIST Center for Neutron Research,
National Institute of Standards and Technology,
Gaithersburg, MD 20899}
\author{S. Ciraci}
\affiliation{Department of Physics, Bilkent University,
Ankara 06533, Turkey}

\date{\today}

\begin{abstract}

We show that the electronic and atomic structure of carbon nanotubes
undergo dramatic changes with hydrogen chemisorption from first
principle calculations. Upon uniform exohydrogenation at half coverage,
the cross sections of zigzag nanotubes become literally square or
rectangular, and they are metallic with very high density of states at
the Fermi level, while other isomers can be insulating. For both zigzag
and armchair nanotubes, hydrogenation of each carbon atom from inside
and outside alternatively yield the most stable isomer with a very weak
curvature dependence and a large band gap.

\end{abstract}



\maketitle

Single wall carbon nanotubes,\cite{iijima,dressel} SWNTs, are among the
most attractive systems for fabricating nanodevices because they exhibit
many unusual mechanical and electronic properties. Variation in the
chiral vector or a small radial deformation result in marked changes
ranging from insulating to ideal one\textendash dimensional (1D)
conducting properties.\cite{hamada,white,cntsts,zyao,oguz1}
The physical and chemical properties SWNTs can also be efficiently
engineered by the adsorption of atoms or molecules on
nanotubes.\cite{dillon,kudin,tada,yuchen,milee,spchan,kudin2,htrans}
Recently, it has been shown that the chemical activity of carbon nanotubes
also depend on the chirality and radius,\cite{tubech,tubestr}
denoting tunable absorption of atoms on SWNTs by structural
deformation.\cite{tubestr,deepak} The interplay between adsorption and
electromechanical properties can give rise to novel physiochemical
properties.\cite{tubestr,deepak} The experimentally observed sensitivity
of the electronic properties of SWNTs to the presence of oxygen and
hydrogen is clear evidence for the importance of this
interplay.\cite{impur}

Motivated by these considerations, we have investigated the structural
and electronic properties of hydrogenated SWNTs (H-SWNT) as a function
of hydrogen coverage and decoration (i.e. isomers) by extensive first
principles calculations. Our results indicate that hydrogen adsorption
on nanotubes gives rise to many novel properties which can mediate
important applications in molecular electronics. One of our most
important results is that upon hydrogenation at uniform half coverage,
the zigzag $(n,0)$ SWNTs are metallized with high density of states at
the Fermi level and the circular cross sections of the tubes are changed
to square or rectangle ones. The carbon atoms near the corners form
new diamond\textendash like C\textendash C bonds. Therefore, these
carbon atoms are electronically and chemically passive, isolating the
four conducting faces of the H-SWNT. Hence, loosely speaking a uniform
half-coverage (n,0) H-SWNT is composed of \textit{four-wire nanocable}.

\begin{figure}
\includegraphics[scale=0.45]{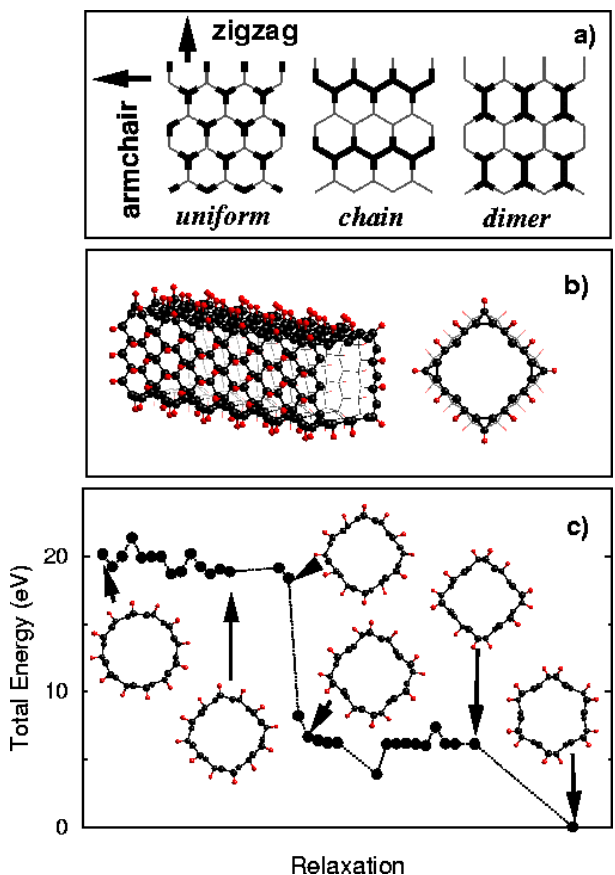}
\caption{
(a) A view of three different isomers of H-SWNT at half coverage.
The left and up arrows indicate the tube axis for armchair and
zigzag nanotubes, respectively. Carbon atoms which are bonded
to hydrogens are indicated by dark color.
(b) A side and top view of a $(12,0)$ H-SWNT, indicating the
square cross section of a uniformly exohydrogenated nanotube
at half coverage.
(c) Several snap shots during the relaxation steps  of an armchair
$(6,6)$ H-SWNT, indicating that an uniform exohydrogenation at half
coverage is not stable against forming a chain isomer.
}
\label{fig:dimer}
\end{figure}

Our study comprises zigzag ($(7,0)$, $(8,0)$, $(9,0)$, $(10,0)$, $(12,0)$)
and armchair ($(6,6)$, $(10,10)$) SWNTs which are hydrogenated at two
different coverages. For full coverage ($\Theta =$1), we consider two
isomers; namely
\textit{(i)} \textbf{exohydrogenation} where each carbon atom is bonded
to a hydrogen atom from outside of the nanotube
(labeled by C$_{4n}$H$_{4n}$) and
\textit{(ii)} \textbf{endo-exohydrogenation} where each carbon atom is
bonded to a hydrogen from inside and outside of the tube alternatively
(labeled by C$_{4n}$H$_{2n}$H$_{2n}$).
For half coverage ($\Theta=$0.5), we consider the three most interesting
isomers  (labeled by C$_{4n}$H$_{2n}$); namely
\textit{(i)} \textbf{uniform} pattern where every other carbon atom is
bonded to a hydrogen from outside,
\textit{(ii)} \textbf{chain} pattern where every other carbon zigzag
chain is saturated by hydrogen, and
\textit{(iii)} \textbf{dimer} pattern where every other carbon dimer
rows perpendicular to the zigzag carbon chains are saturated by hydrogen.
Fig.~\ref{fig:dimer}a shows these three isomers at half coverage.

The first principles total energy and electronic structure calculations
have been carried out within the generalized gradient approximation
(GGA)~\cite{gga} using the pseudopotential plane wave method~\cite{castep} 
in a supercell geometry.
Details of the parameters used in
this work are the same with those given in Ref.\onlinecite{tubestr}.
Fully relaxed geometries are obtained by optimizing all atomic positions
and the lattice constant $c$ along the tube-axis until the maximum force
and stress are less than 0.01~eV/\AA{} and 0.1~GPa, respectively.

\begin{figure}
\includegraphics[scale=0.46]{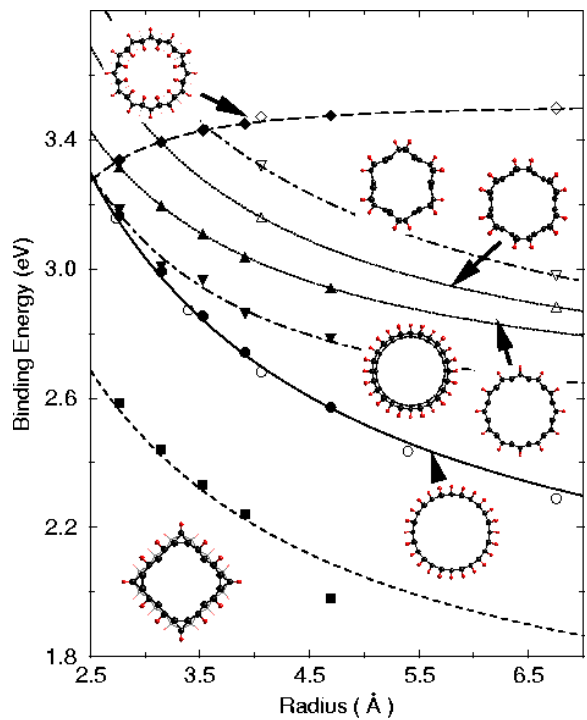}
\caption{
Average binding energies, $E_b$, of hydrogen atoms adsorbed on various
zigzag and armchair SWNTs versus bare tube radius $R$. Filled and open
symbols are for zigzag and armchair nanotubes, respectively. Circles
and diamonds are for exo- and endo-exohydrogenation at full coverage,
respectively. The filled squares show the zigzag nanotubes uniformly
exohydrogenated at half coverage. The chain and dimer patterns of
adsorbed hydrogen atoms at half coverage are shown by down- and
up-triangles, respectively. Curves are analytical fits explained
in the text. Insets show top view of several H-SWNT isomers.}
\label{fig:binding}
\end{figure}

We find that geometric and electronic structures and binding energies
of H-SWNTs strongly depend on the pattern of hydrogenation
(i.e. decoration). The most remarkable effect is obtained when zigzag
nanotubes are uniformly exohydrogenated at half-coverage ($\Theta=$0.5).
Upon hydrogenation the structure undergoes a massive reconstruction,
whereby circular cross section of the $(7,0)$ SWNT changes to a rectangular
one, and those of $(8,0)$, $(9,0)$, $(10,0)$ and $(12,0)$ change to square
ones as shown in Fig.~\ref{fig:dimer}b. These new structures are stabilized
by the formation of new diamond\textendash like C\textendash C bonds with
$d_{CC} \sim$ 1.51-163 \AA~ near the corners of rectangular or square
H-SWNTs.
Hence, triangular and pentagonal C rings are formed instead of hexagonal.
Depending on ($2n \bmod 4$), either one bond is formed just at the corners
or two bonds at either side of the corners. Most interestingly, all these
structures are found to be metallic with a \textit{very large density of
states at the Fermi level.} The uniform adsorption at $\Theta=$0.5 for
zigzag nanotubes are metastable. Such a local minimum does not exist for
armchair nanotubes, since uniformly adsorbed H atoms are rearranged upon
relaxation by concerted exchange of C\textendash H bonds to form zigzag
chains along the tube axis. Several snap shots for hydrogen dimerization
on an armchair tube are shown in Fig.~\ref{fig:dimer}c.
The cross sections of chain isomer at $\Theta=0.5$ of armchair tubes are
polygonal where the corners are pinned by the zigzag H chains along the
tube axis. For other isomers at half-coverage as well as exo- and
endo-exohydrogenations at full coverage, the cross sections remain
quasi-circular (see Fig.~\ref{fig:binding}).

The average binding energies of H are obtained from the expression,
\begin{equation}
E_b= (E_{T,C_{4n}}+ m E_H - E_{T,C_{4n}H_{m}})/m
\label{eq:binde}
\end{equation}
in terms of the total energies of the bare SWNT, $E_{T,C_{4n}}$, the
hydrogen covered SWNT, $E_{T,C_{4n}H_{m}}$, and the energy of atomic
hydrogen, $E_H$ ($m$ is the number of H atoms per unit cell). According
to the above definition stable structures have positive binding energies.
The average binding energies as a function of nanotube radius are shown
in Fig.~\ref{fig:binding}. The variation of $E_b$ with the radius of the
bare nanotube can be fitted to a simple formula,\cite{tubech,tubestr}
\begin{equation}
E_b = E_{o}(\Theta) + \frac{C_{p}(\Theta)}{R^p}.
\label{eq:fit}
\end{equation}
Note that $1/R$ form (i.e. $p=1$) is quite common to SWNTs and scales
various properties.\cite{white,cntsts,oguz1,tubech,tubestr} Here
$E_{o}(\Theta)$ is the binding energy when $R \rightarrow \infty$, and
hence corresponds to the adsorption of H on graphene at a given coverage,
while $C_{p}(\Theta)$ is a constant that depends on coverage $\Theta$,
and represents the curvature effect. Calculated $E_b$'s are fitted to
Eq.~\ref{eq:fit} with the values listed in Table~\ref{table:fit}.
While $E_b$ increases with decreasing $R$ in the case of exohydrogenation,
this trend is reversed for endo-exohydrogenation due to increased
H\textendash H repulsion inside the tube at small $R$. Nevertheless, the
endo-exohydrogenation of SWNTs, which transforms the $sp^2$ to $sp^3$-like
bonding, gives rise to the highest binding energy saturating at 3.51 eV
as $R\rightarrow \infty$. At this limit, the exo-endo hydrogenated
graphene (from above and below) is buckled by 0.46 \AA~ as if two diamond
(111) planes with interplanar distance of 0.50 \AA~.

\begin{table}
\caption{Values of the parameters $E_{0}$(eV) and
$C_{p}(\Theta)$ (eV \AA$^{p}$) given in Eq.(\protect\ref{eq:fit})
to fit the binding energies.}
\label{table:fit}
\begin{ruledtabular}
\begin{tabular}{l||l|l|l}
 Hydrogen   &            & Zigzag Tubes     & Armchair Tubes \\
decoration  &  $E_{0}$   & $C_{p}(\Theta)$  & $C_{p}(\Theta)$ \\ \hline
Exo $\Theta = 1.0$      & 1.75 & $C_{1}=3.87$  & $C_{1}=3.87$ \\
Exo-endo $\Theta = 1.0$ & 3.51 & $C_{3}=-3.49$ & $C_{3}=-3.49$ \\
Uniform $\Theta = 0.5$  & 1.41 & $C_{1}=3.19$  & not stable \\
Chain   $\Theta = 0.5$  & 2.50 & $C_{1}=0.53$  & $C_{1}=3.16$ \\
                        &      & $+C_{2}=3.70$ & $+C_{2}=0.73$ \\
Dimer   $\Theta = 0.5$  & 2.55 & $C_{1}=1.41$  & $C_{1}=1.87$ \\
                        &      & $+C_{2}=1.94$ & $+C_{2}=2.42$ \\
\end{tabular}
\end{ruledtabular}
\end{table}

\begin{figure}
\includegraphics[scale=0.4,angle=-90]{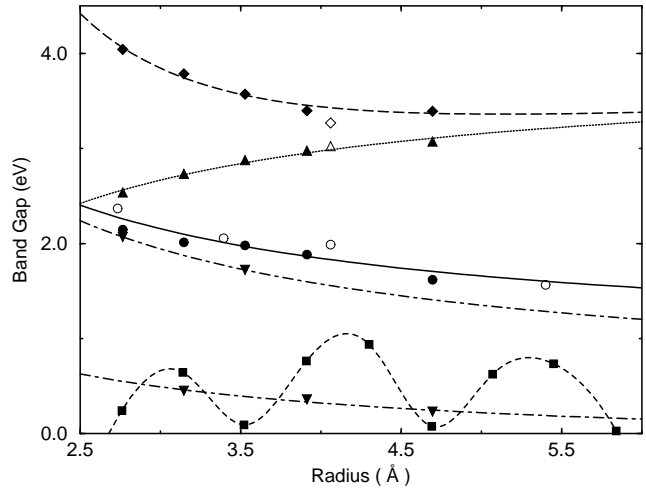}
\caption{
The band gaps, $E_g$, versus the bare nanotube radius $R$.
Filled and empty symbols indicate zigzag and armchair SWNTs, respectively.
Squares show non-monotonic variation of the band gap of the bare zigzag
nanotubes. Exo and endo-exohydrogenated nanotubes ($\Theta=1$) and
chain and row patterns of adsorbed hydrogen atoms at half coverage
are shown by circles, diamonds, down- and up-triangles, respectively.
Lines are guide to eye.}
\label{fig:bgap}
\end{figure}

As with the atomic structure, the electronic structure of SWNTs undergo
important changes as a result of hydrogenation. For bare zigzag SWNTs
the variation of the band gap, $E_g$, with $n$ is rather complex due to
interplay between zone folding and curvature induced $\sigma^* - \pi^*$
mixing\cite{oguz1,blase} while bare armchair SWNTs are metallic. Here
we find that sizable band gaps are opened as a result of hydrogenation
(Fig.~\ref{fig:bgap}). At $\Theta=$1, the band gap displays a similar
behavior for both type of nanotubes and hydrogenation, and decreases with
inreasing $R$. Relatively larger band gaps (in the range of 3.5-4 eV) of
endo-exohydrogenated SWNTs ($\Theta=$1) can be explained by the fact that
the adsorption of H alternatively inside and outside leads to the atomic
configuration closer to the diamond structure having rather large band gap
($E_g=$5.4 eV).

The effect of hydrogenation on the electronic structure is even more
remarkable at $\Theta = $ 0.5 and is the most interesting aspect of our
study. Depending on the pattern of hydrogen adsorption, an isomer can be
either a metal or insulator. For example, all uniform (n,0) H-SWNTs are
metallic. On the other hand, the chain pattern realized on the $(n,0)$
SWNTs results in two doubly degenerate, almost dispersionless states at
the valence and conduction band edges. The band gap $E_g$ between these
states decreases with increasing $R$. When $n$ is odd, $E_g$ is large
(e.g $E_g =$2.1 eV for $(7,0)$). When $n$ is even the doubly degenerate
band at the conduction band edge moves towards the valence band edge
and splits as bonding and anti-bonding states. As a result $E_g$ is
reduced significantly becoming only a pseudogap for large and even $n$.
Finally, the dimer row isomers are insulators and the $E_g$ increases
with increasing radius. Surprisingly, there are two dispersive bands
with $\sim 1$~eV bandwidth at both band edges and extremum moves from
the center of Brillioun zone ($\Gamma$-point) to the zone edge ($Z$-point).

\begin{figure}
\includegraphics[scale=0.49]{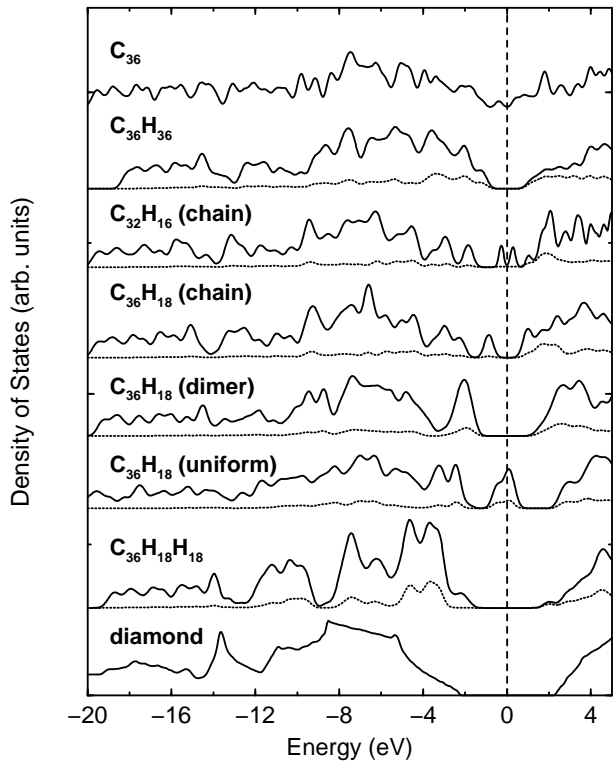}
\caption{Comparison of the electronic density of states
(DOS) of a bare $(9,0)$ nanotube (C$_{36}$) and its various hydrogenated
isomers. DOS of a chain pattern (8,0) H-SWNT at half coverage
($C_{32}H_{16}$) and bulk diamond (bottom panel) are also shown for
comparison. The zero of energy is taken at the Fermi energy showed
by vertical dashed line. The dotted lines are the partial density of
states resolved on hydrogen atoms.}
\label{fig:dos}
\end{figure}
 
Hydrogen adsorption induced dramatic changes of electronic structure are
demonstrated by total density of states (DOS), $\mathcal{D}(E)$, of
$(9,0)$ nanotubes in Fig.~\ref{fig:dos}. First of all, the small band gap
of the bare $(9,0)$ SWNT is opened by $\sim 2$~eV upon exohydrogenation
at $\Theta=$1. The band gap is still significant for $\Theta=0.5$ with
the chain pattern, and increases to 4 eV for dimer pattern. However,
a similar chain pattern in the $(8,0)$ H-SWNT (see $C_{32}H_{16}$ in
Fig.~\ref{fig:dos}) has a much smaller band gap. Suprisingly, all zigzag
nanotubes uniformly exohydrogenated at $\Theta=$0.5 are metals.
As displayed in the sixth panel of Fig.~\ref{fig:dos} for
$C_{36}H_{18}$(uniform), their total density of states are characterized
by a peak yielding high state density at $E_F$. While carbon states are
pushed apart, yielding a $\sim$ 4-5~eV gap, a new dispersive metallic
band with $\sim$ 1-2~eV bandwidth crosses the Fermi level. Apart from
being an ideal 1D conductor, this \textit{very high density of states
at $E_F$ might lead to superconductivity.} Note that, these uniform H-SWNTs
undergo a massive reconstruction and their circular cross sections change
into square ones with the formation of new C\textendash C bonds ($C_4$)
at the corners. All C atoms without H attached (except those at corners)
as well as the H atoms at the center of four planar sides contribute to
the high $\mathcal{D}(E_F)$. This way four individual conduction paths
are formed on each side of square tube. It is emphasized that the
transformation from the $sp^2$- to the $sp^3$ bonding underlies various
effects discussed in this study. The H-SWNTs, especially
C$_{4n}$H$_{2n}$H$_{2n}$ structures can be conceived as if they are more
diamond-like than graphitic. Our arguments are justified by the comparison
of $\mathcal{D}(E)$ of the endo-exohydrogenated $(9,0)$ with that of bulk
diamond in Fig.~\ref{fig:dos}. Apart from opening a large band gap,
the quasi metallic $\mathcal{D}(E)$ of the bare $(9,0)$ is modified to
become similar to that bulk diamond. The latter has relatively larger
valence band width due to coupling of distant neighbors.

In conclusion, our study reveals many important and novel effects of
hydrogen adsorption on SWNTs, and brings a number of new problems and
issues to be explored. For example, one can argue that the band gap of
a SWNT can be engineered by the controlled hydrogenation of a single
nanotube as in the alloy of Si$_{x}$Ge$_{1-x}$. A number of isomers which
can be tailored with different hydrogen decoration provide options in
developing new materials. Furthermore, multiple quantum well structures,
or one dimensional chain of quantum dots, can be tailored by periodic and
modulated hydrogenation of a single nanotube.
Finally, \textit{the very high density of states at the Fermi level} of
uniform pattern isomer at half coverage may result in to
\textit{superconductivity} in SWNT based nanowires. Needless to say,
realization of the systems proposed here will be an experimental challenge.
However, the fact that other carbon clusters such as cubane, dodecahedrane,
and C$_{60}$H$_{32}$ have been successfully synthesized suggests that
this is not impossible.
 
\textbf{Acknowledgments}
This work was partially supported by the NSF under Grant No. INT01-15021
and T\"{U}B\'{I}TAK under Grant No. TBAG-U/13(101T010).

\end{document}